\documentstyle[twocolumn,aps,epsfig]{revtex}
\tightenlines
\begin{document}
\twocolumn[\hsize\textwidth\columnwidth\hsize\csname @twocolumnfalse\endcsname
\title{ZrZn$_2$: geometrical enhancement of the local DOS and quantum
design of magnetic instabilities}
\author{Ezio Bruno$^\dag$\cite{ezio}, Beniamino Ginatempo$^\dag$ and J.B. Staunton$^\ddag$}

\address{\dag Dipartimento di Fisica and Unit{\`{a}} INFM,
Universit{\`{a}} di Messina, Salita Sperone 31, 98166 Messina, Italy}
\address{\ddag Physics Department, University of Warwick, UK}
\date{\today}
\maketitle

\begin{abstract}
The recent discovery of coexisting ferromagnetism and superconductivity
in $ZrZn_2$, and the fact that they are simultaneously suppressed on
applying pressure\cite{SMHayden} suggest the possibility of a pairing
mechanism which is
mediated by exchange interactions and connected with the proximity
to a magnetic quantum critical point. On the basis of first principles,
full potential electronic structure
calculations, we study the conditions that, for $ZrZn_2$, determine the
proximity to this magnetic instability. More specifically, we discuss the role
played by the geometrical arrangement of the lattice, hybridization  effects and the
presence of disorder,  as well as the  application of external
pressure. These circumstances influence the width of the relevant Zr d
bands whose narrowing, due to the reduction of the {\it effective} number
of neighbours or to an increase of the cell volume, causes an {\it enhancement}
of the density of states at the Fermi level. Finally, we highlight some general features
that may aid the design of other materials close to
magnetic instabilities.

\end{abstract}

\newpage
]
The very recent discovery of the coexistence of ferromagnetism (FM) and
superconductivity (SC) in $ZrZn_2$\cite{SMHayden} and $UGe_2$ \cite{Saxena}
suggests that the current ideas on SC need to be critically reconsidered.
Interestingly $ZrZn_2$ is a  relatively 'simple' metal, in the sense that
there are not complications arising from $f$ electrons \cite{lda+u} or oxide
planes. This circumstance implies that its electronic properties can
be studied in detail within the density functional theory\cite{HK_KS} framework
which is expected to be reliable for such systems.

As known for many years, $ZrZn_2$ is a weak itinerant
FM~\cite{Matthias&Bozorth}
with a Curie temperature $T_{FM}=28.5 K$, despite the fact
that both pure Zr and Zn are not magnetic. Both pure elements also exhibit
superconducting transitions with
$T_C=0.6 K$ and $T_C=0.85 K$, respectively~\cite{CRC}. Many speculations on
the properties of this compound have been stimulated by the
discovery~\cite{Smith} that, at some critical pressure, $P_C \simeq 8
KBar$, it becomes
paramagnetic (PM), suggesting that it is very close to a magnetic quantum
critical point (MQCP). Notwithstanding the argument of Berk and
Schrieffer~\cite{Berk&Schrieffer} about the suppression of the phonon-induced
s-wave SC by FM spin fluctuations,
many authors~\cite{many} argued that, in the vicinity of the MQCP
on the PM side (high pressures), p-wave SC could
be induced by paramagnons. Later on, Fay and Appel~\cite{Fay}
suggested the possibility of having p-wave SC also for
$P<P_C$, {\it inside} the FM region of the phase diagram. A different scenario,
recently proposed by Blagoev et al.~\cite{Blagoev}, predicts s-wave
SC for $P<P_C$ and p-wave SC for $P>P_C$.

The recent experiments~\cite{SMHayden} suggest that the kind
of SC observed in $ZrZn_2$ is related to the presence of FM, since both
FM and SC are suppressed simultaneously as the pressure reaches $P_C$.
This new phenomenon could be compatible either with the picture of
Ref.~\cite{Fay} or with that of Ref.~\cite{Blagoev}, both assuming a magnetic
excitation-mediated coupling in the proximity of the MQCP. In order to
decide between alternative explanations it is highly desirable to be
able to {\it engineer} materials close to a MQCP and to control the
vicinity by changing some variable such as the external
pressure or the concentration of some impurities. In this letter, on the basis
of first-principles, full-potential, electronic structure calculations
for both $ZrZn_2$ and related systems, we highlight a rather general feature
which may aid this endeavor.

In particular we examine the $ZrZn_2$
lattice geometry and follow some important consequences for the physical
properties.
We find that the density of states (DOS) at
Fermi energy, $n(E_F)$ is {\it enhanced}, in a sense to be specified
below, by the lattice {\it geometry}. A high $n(E_F)$ is a crucial attribute for
a material to be close to an MQCP as shown, for example, by a simple
Stoner factor calculation
(a measure of proximity to a ferromagnetic
phase transition)
\begin{equation}
S_0=\left ( 1 -I n(E_F)\right)^{-1}
\label{Stoner},
\end{equation}
where $I$ is the exchange integral.
Of course, a large $n(E_F)$ is also important in generating a sizeable
electron-phonon coupling,
\begin{equation}
\lambda=\frac{n(E_F)<I^2>}{M<\omega^2>}
\label{e-phon}.
\end{equation}
where $<I^2>$ is the electronic stiffness parameter of Gaspari and Gyorffy
\cite{Gaspari&Gyorffy}, M the ionic mass and $<\omega^2>$ some mean
value of the phonon frequency. In the above equations we have deliberately
used definitions from the simplest available electronic theories for magnetism
and SC, just to stress the relevant role played in these phenomena by $n(E_F)$.

\begin{figure}
\centerline{\epsfig{figure=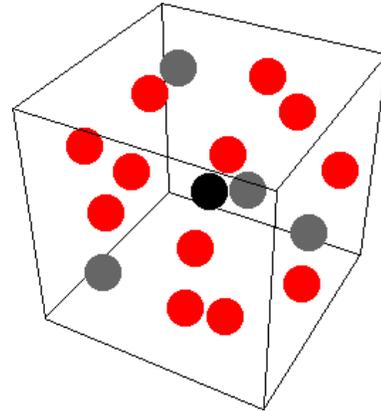,width=5cm}}
\caption{The local environment of a Zr atom (marked in black) in the C15 cubic
Laves structure. Other Zr atoms: dark circles; Zn: red circles}
\label{geometry}
\end{figure}

$ZrZn_2$ crystallizes into a cubic C15 superlattice~\cite{weblattices}.
In this structure, Zr atoms occupy the positions of a diamond lattice
while the Zn atoms form a network of interconnected tetrahedra. Since the
major contributions to $n(E_F)$, as we shall see, come from Zr, the local
environment of Zr atoms (Fig.~\ref{geometry}) is particularly important for our
concerns. Each Zr is surrounded by 12 Zn neighbours and 4 Zr neighbours,
at distances that in terms of the lattice constant, $a$, are, respectively,
$d_{ZrZn}=\frac{\sqrt{11}}{8} a$ and $d_{ZrZr}=\frac{\sqrt{3}}{4} a$. In
other words, the distance between two Zr atoms is only 4 per cent larger
than the nearest neighbours (NN) distance $d_{ZrZn}$. This circumstance leads to
an appreciable overlap of the wave-functions of two neighbouring Zr
atoms~\cite{Jarlborg&Freeman}.

We have calculated the band structure of $ZrZn_2$ in the PM state using the
full potential linearised augmented plane waves method\cite{WIEN97} (FLAPW)
within the local density approximation (LDA). The calculated equilibrium
lattice constant (13.58 a.u.) is about 3 percent smaller than the experimental
one (13.98 a.u.~\cite{SMHayden}). Importantly, our full potential results
provide a reference for more approximate,
existing band structure
calculations~\cite{Jarlborg&Freeman,Morozumi}, all based on the linearized
muffin-tin orbital (LMTO) method and the atomic sphere approximation (ASA).
These simpler band structure calculations form the starting point for studies
of electron-phonon coupling and spin fluctuations in this material.

\begin{figure}
\centerline{\epsfig{figure=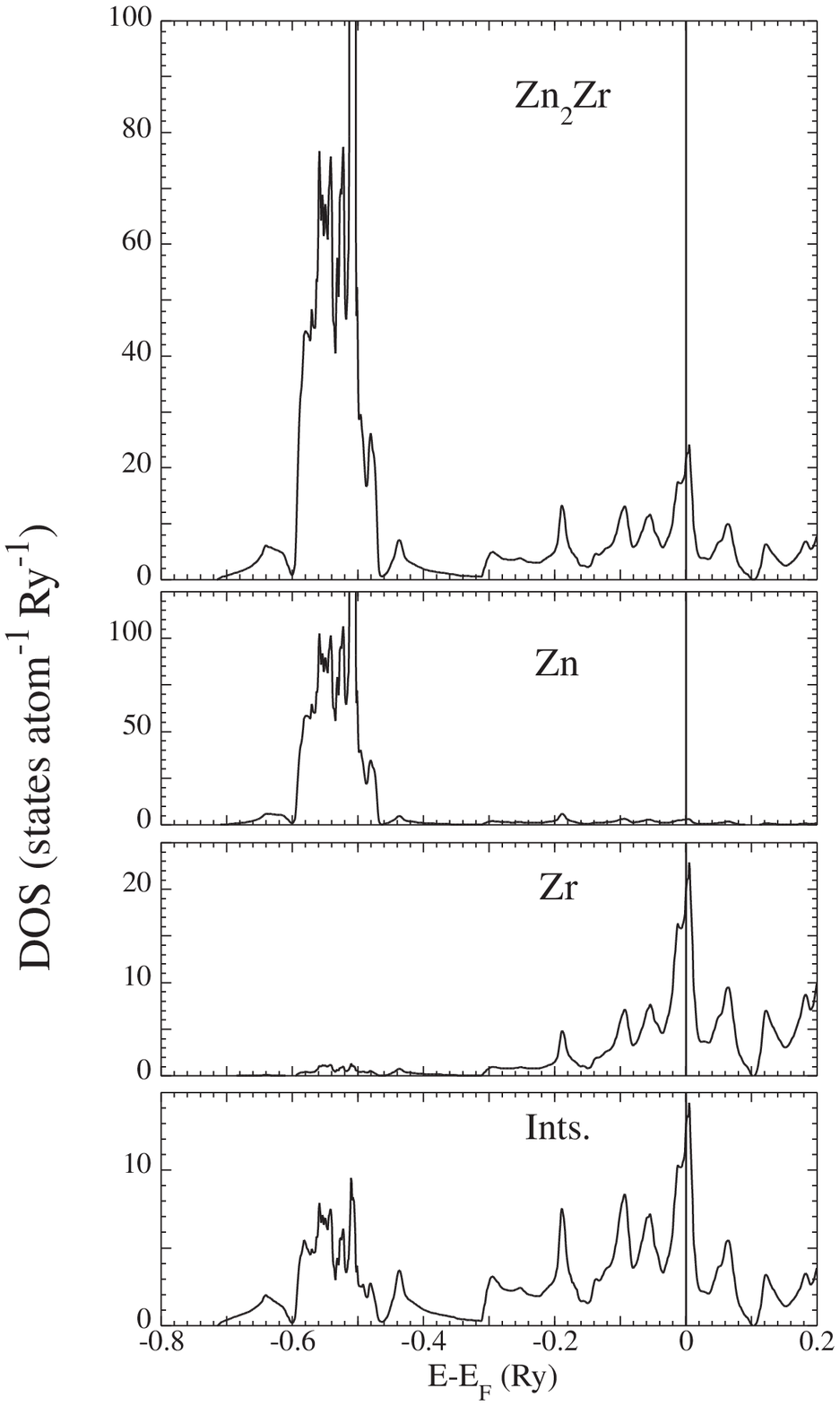,width=7cm}}
\caption{DOS of PM $ZrZn_2$ (top panel) at the relaxed lattice constant,
a=13.58 a.u.. The labels Zn and Zr
refer to the corresponding muffin-tin spheres. The lower panel shows the
interstitial DOS divided by the number of atoms in the unit cell.}
\label{fulldos}
\end{figure}

\begin{figure}
\epsfig{figure=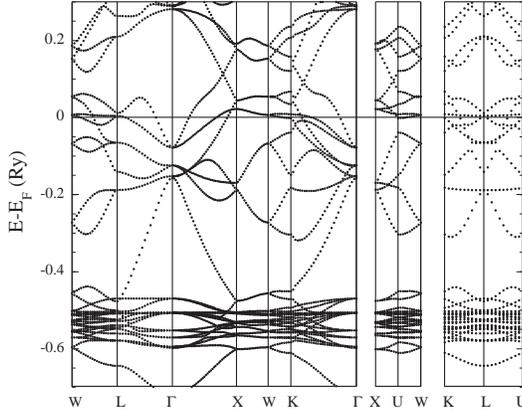,width=7cm}
\caption{Band structure of PM $ZrZn_2$ along relevant symmetry directions.}
\label{bands}
\end{figure}

\begin{figure}
\centerline{\epsfig{figure=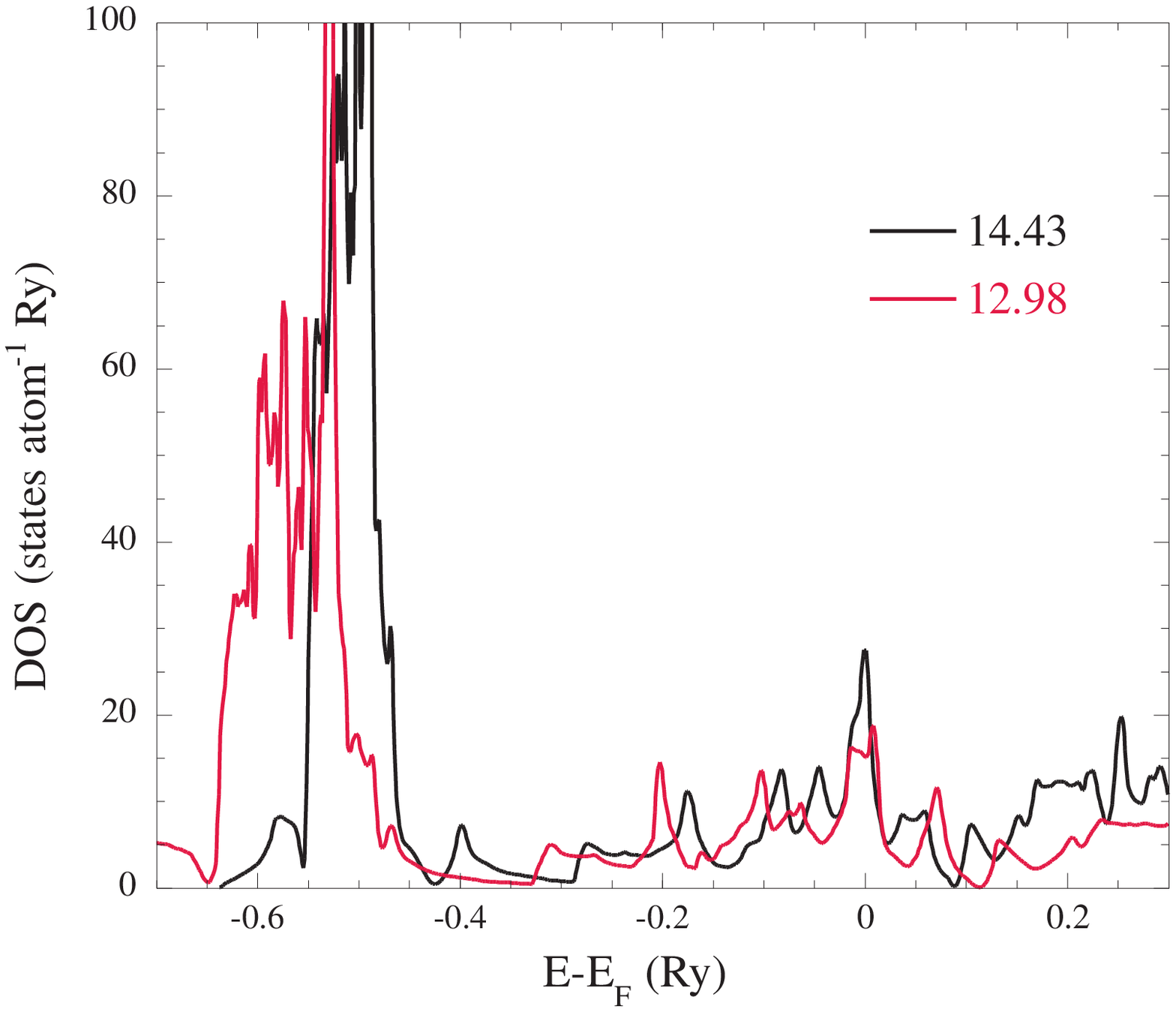,width=7cm}}
\caption{DOS of PM $ZrZn_2$ for the values of $a$ indicated by the labels}
\label{narrowing}
\end{figure}

\begin{figure}
\centerline{\epsfig{figure=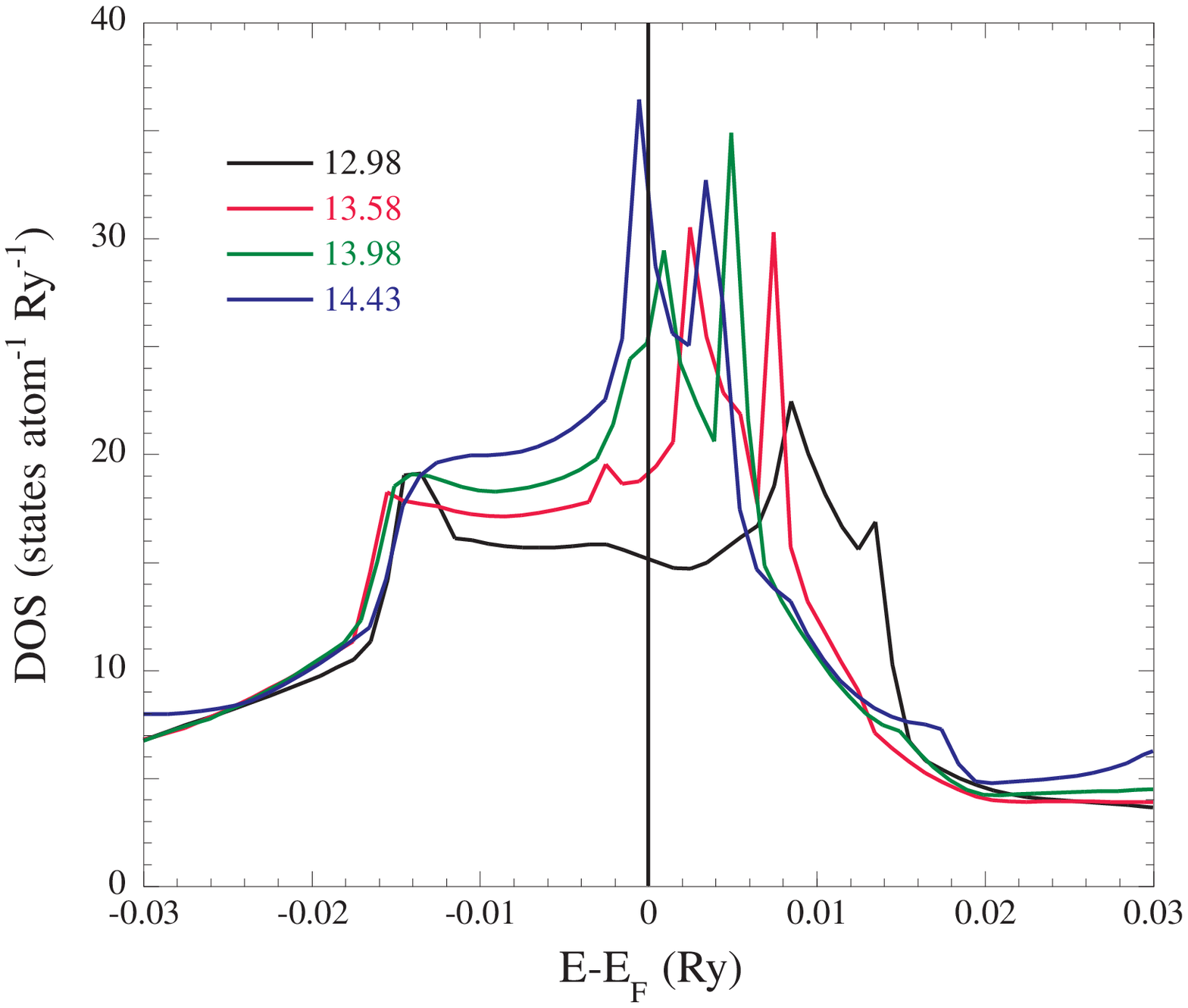,width=7cm}}
\caption{DOS of PM $ZrZn_2$ about $E_F$ for the values of $a$
indicated by the labels.}
\label{pressure}
\end{figure}

The DOS (Fig.~\ref{fulldos}) classifies this
compound as a {\it split band} system: the electronic states are distributed
into largely separate energy regions. The Zr d resonances are close
to $E_F$, while those of Zn occur well below. The band structure, plotted in
Fig.~\ref{bands}, reveals very narrow d bands and suggests
that these would be appreciably broadened out by a reduction of the interatomic
distances. A similar broadening could also be caused by
decrease of the quasiparticle lifetimes due to impurities or defects.
This view is supported by Fig.~\ref{narrowing}, where we see how the
increase of the bandwidth for smaller values of $a$ results in a dramatic drop
of the DOS peaks heights. For the system at hand, the drop of $n(E_F)$ is also
affected by electronic topological transitions~\cite{ETT} (ETT) occurring
when the
Fermi level crosses the various DOS peaks, as shown in Fig.~\ref{pressure}.

\begin{figure}
\centerline{\epsfig{figure=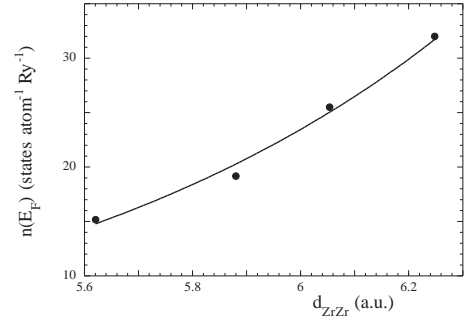,width=6cm}}
\caption{$n(E_F)$ vs. $d_{ZrZr}$ in $ZrZn_2$. The circles correspond to the
calculated values, the line is an exponential fit. We find B=1.2 (see the
text), in good agreement with the values reported in Ref.[18].}
\label{nef}
\end{figure}

The variation vs. $a$ of the local DOS in Zr muffin-tin spheres at the Fermi level,
$n_{Zr}(E_F)$, is displayed in Fig.~\ref{nef}. These trends can be
easily fitted using a simple tight-binding (TB) model~\cite{papa}: $n_{Zr}(E_F)$
is inversely proportional to the bandwidth which is proportional to the
hopping parameters. These scale exponentially with the relevant
distances, say as $e^{-Br}$ with $B>0$.
Accordingly with Eq.~\ref{Stoner}, the behaviour of $n_{Zr}(E_F)$ is consistent
with the
experimental findings~\cite{SMHayden,Smith} that a compression can suppress
the magnetic ordering by crossing, at $P=P_C$, the
MQCP ($S_0^{-1}=0$). On the other hand, consistently with
Ref.~\cite{SMHayden}, the MQCP could be reached at lower pressures in
impure samples, due to the lifetimes effects mentioned above.

Pure Zr forms a hcp crystal structure. Although very well
known, the standard explanation as to why it is not ferromagnetic
contains some valuable pointers.
$n_{Zr}(E_F)=4.11$ is about 4 times smaller than in $ZrZn_2$. Such a
large factor cannot be simply attributed to the atomic distances
involved (see Fig. ~\ref{nef}). For pure (hcp)
Zr we find $d_{ZrZr}=5.900$ a.u., very close to both $d_{ZrZr}$ and
$d_{ZrZn}$ in $ZrZn_2$. Neither can this factor
be caused entirely by ETT's. Such a decrease can be explained only
by assuming that, as a consequence of the very weak hybridization with Zn,
Zr atoms 'see' only their 4 Zr neighbours, while in pure Zr the
number of NN's, $f$, is 12. This could account for the
reduction of the local DOS by factor 3. Using the above TB model, we can draw up
a relationship
\begin{equation}
n_{Zr}(E_F) \sim A e^{B d_{ZrZr}}/f_{eff}
\label{TB},
\end{equation}
where $f_{eff}$ is some effective number of NN's. Of course, Eq.~\ref{TB}
should not be taken too literally since ETT's and differences between the
shapes of the DOS's in different lattices (e.g. the pseudogap between
bonding and antibonding states in hcp lattices) have also some
influence.

\begin{figure}
\centerline{\epsfig{figure=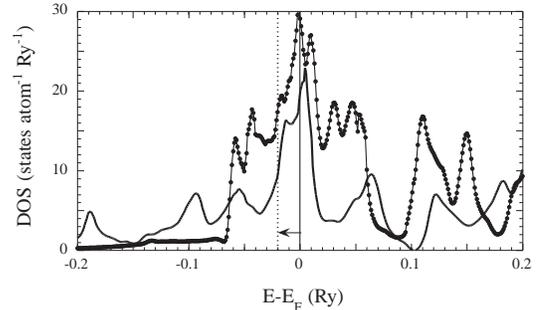,width=7cm}}
\caption{The density of states of pure Zr in the diamond lattice (dots) compared
with the density of states of $ZrZn_2$ (plain line), at the relaxed lattice
constant of $ZrZn_2$.
All the DOS refer to Zr muffin tin spheres only. The arrow indicates the
direction in which the Fermi level is espected to move due to charge transfer
from Zr to Zn. The dashed line marks an upper bound estimate of the above Fermi
level shift.}
\label{diamond}
\end{figure}

In order to add substance to our picture of $ZrZn_2$ in which the Zn atoms
play essentially the role
of empty spheres placed between Zr atoms to reduce $f_{eff}$, we
have calculated the bandstructure of Zr in
the hypothetical diamond lattice phase that would result by the
substitution of the Zn atoms in the
$ZrZn_2$ lattice with empty spaces. The comparison of the corresponding
DOS (Fig.~\ref{diamond}) shows up some minor differences mostly due
to the weak hybridisation with Zn and with the charge transfer to Zn atoms in
the $ZrZn_2$ system.

In summary: we have pointed out how the addition of Zn to an ideal Zr diamond
lattice changes the local DOS on Zr atoms very little. We surmise that
this is an example of a rather common occurrence. One can consider a
hypothetical geometrical lattice structure formed by adding empty spaces
in some metal. The narrowed relevant bands could give rise to exotic, magnetic
or superconducting, properties (as shown here for Zr in the diamond lattice,
that, of course, cannot be realized in practice). Those same properties could
then be obtained by filling the empty spaces with atoms which give rise to
little (or zero) DOS at the Fermi level (as Zn) and which have little
hybridization with the metal we want to modify. Contrary to
the fashion prevalent in much scientific literature, we close this
Letter by saying that the observations we are making from our calculations are
not particularly new. The focus, however, on the geometrical guideline we have
extracted is. A list of existing systems whose properties seem to be
consistent with this guideline can be easily constructed. For example, there are
thin 4d and 5d metal films~\cite{Pd}, where the reduction of $f_{eff}$ at the
surface allows magnetically ordered states to be stable. Another example might
be the alkali fullerides~\cite{fullerides}, where $C_{60}$ molecules could
provide the 'vacuum' making alkali metals superconducting. Other possible
examples are $MgB_2$~\cite{Akimitsu}, $UGe_2$~\cite{Saxena},
$MnSi$~\cite{Fay,MnSi} the layered cuprate
~\cite{hightc} and ruthenate superconductors~\cite{Maeno} and $A15$
compounds~\cite{A15}.

{\it Acknowledgements:}   We are grateful to E.S. Giuliano, B.L. Gyorffy and
G. Santi for helpful discussions.

\newpage

\end{document}